\documentclass[journal,twoside]{IEEEtran}
\usepackage[LGR,T1]{fontenc}
\usepackage{amsmath,amsfonts}
\usepackage{algorithmic}
\usepackage{algorithm}
\usepackage{array}
\usepackage[caption=false,font=normalsize,labelfont=sf,textfont=sf]{subfig}
\usepackage{textcomp}
\usepackage{stfloats}
\usepackage{url}
\usepackage{verbatim}
\usepackage{graphicx}
\usepackage{cite}
\usepackage{xcolor}
\usepackage[detect-all]{siunitx}
\usepackage{adjustbox}
\usepackage[para]{threeparttable}
\usepackage{multirow}
\usepackage{footnote}
\usepackage{booktabs}
\usepackage{pifont}
\usepackage{lipsum}
\usepackage{microtype}

\newcommand{\firstRev}[1]{\textcolor{black}{#1}}
\newcommand{\genRev}[1]{\textcolor{black}{#1}}

\begin{document}
\bstctlcite{IEEEexample:BSTcontrol}

\title{\firstRev{A LoRa-Compatible Receiver\\ with Programmable Noise-Power Trade-Off\\ and Integrated Wake-Up Mode}}

\author{Marco~Gonzalez, Pol~Maistriaux, Jérôme~Louveaux and David~Bol
\thanks{M. Gonzalez was with the ICTEAM institute, Université catholique de Louvain, 1348 Louvain-la-Neuve, Belgium. He is now with e-peas semiconductors, 1348 Louvain-la-Neuve, Belgium (e-mail: marco.gonzalez@e-peas.com).
P. Maistriaux, J. Louveaux, and D. Bol are with the ICTEAM institute, Université catholique de Louvain, 1348 Louvain-la-Neuve, Belgium (e-mails: \{pol.maistriaux, jerome.louveaux, david.bol\}@uclouvain.be). \textit{(Corresponding author: Marco Gonzalez)}}%
}

\markboth{}{Gonzalez \MakeLowercase{\textit{et al.}}: A Programmable Wake-Up-Capable LoRa Receiver}

\maketitle

\begin{abstract}
Programmable receivers allow an optimal performance in the variable conditions of the wireless channel. An extensive analysis of their noise-power trade-off enables the development of a main receiver with embedded wake-up circuitry. This makes the seamless integration of wake-up receivers~(WuRXs) in RF receivers possible, avoiding power, area, complexity, and cost overheads in the system.
\firstRev{In this paper, we propose the first LoRa-compatible receiver with an integrated wake-up mode and a programmable noise-power trade-off}. In main-receiver mode, it consumes \qty{4.57}{mW} for a sensitivity of \qty{-138}{dBm}, but it can save up to \qty{62.6}{\%} of power consumption at the cost of an 8.3-dB sensitivity degradation. In wake-up mode, sensitivity down to \qty{-126}{dBm} is achieved for a 32-ms latency and \qty{1.5}{mW} of power consumption. \firstRev{The power consumption remains relatively high due to a sub-optimal design of the the phase-locked loop (PLL). In the end, the wake-up mode consumes 3$\times$ less} than the lowest-noise configuration in main-receiver mode. 
\end{abstract}


\section{Introduction}


\IEEEPARstart{D}{espite} outstanding advances in energy efficiency in Internet-of-Things~(IoT) devices, wireless communication remains one of the most significant contributors to their overall power consumption. This is particularly true for low-power wide-area networks (LPWANs) used in applications like environmental monitoring, where devices can accommodate a low data rate but need a long communication range~\cite{callebaut2021}. Mesh networks and applications with frequent downlink communications significantly increase the pressure on the receiver of the IoT nodes \cite{lee_monitoring_2018,leenders_energy-efficient_2023}. To decrease their average power consumption, they rely on the duty-cycling of their RF receivers, which are enabled in predetermined time slots to demodulate possible incoming packets. Real-time clocks are used to adhere to this schedule, but their inaccuracy and frequency drifts limit the level of duty-cycling that can be implemented in practice, which increases the device's power consumption. Additional synchronization packets are exchanged in networks to compensate for drifts in real-time clocks, which further increases their power consumption \cite{oller_has_2016}.

\begin{figure}[t]
\centering
\includegraphics[width=\columnwidth]{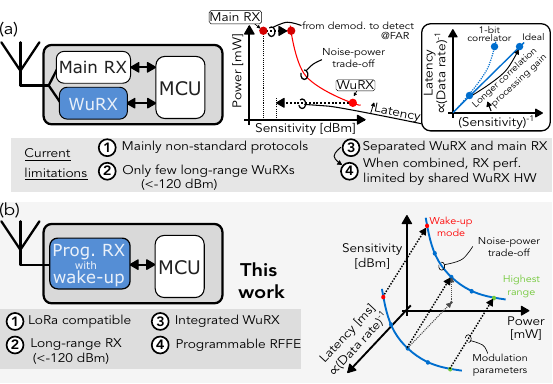}
\caption{(a) General architecture of an IoT device embedding a wake-up receiver, with the sensitivity analysis depicted. Current limitations of the state-of-the-art are presented. (b) Proposed architecture with the programmable LoRa receiver with integrated wake-up capability. The contributions are highlighted.}
\label{fig: intro}
\end{figure}

Wake-up receivers~(WuRXs) have been studied as an alternative to network synchronization. They are ultra-low-power receivers that continuously scan the wireless channel to detect an incoming packet and enable the main receiver for its demodulation. An IoT device embedding a WuRX is depicted in Fig.~\ref{fig: intro}(a). WuRXs need to achieve, at the same time, an ultra-low power consumption and a sufficient sensitivity so that their missed-detection rate (MDR) does not impair the overall performance of the receiver~\cite{kim2021}\cite{mercier_low-power_2022}. 
As analog circuits often exhibit a trade-off between power consumption and noise, WuRXs generally reduce their power by sacrificing their noise figure~(NF)~\cite{gonzalez2024}. To compensate for this, the sensitivity-latency trade-off of receivers is exploited by leveraging a long wake-up signal (WuS) to provide sufficient processing gain through correlations~\cite{kim2023,maistriaux2024}. This way, WuRXs consume less power while maintaining their sensitivity, at the cost of a higher latency for the system, as illustrated in Fig.~\ref{fig: intro}(a).

Several limitations currently prevent the integration of WuRXs in LPWANs. First, non-standard modulation schemes are often used to simplify the WuRX architecture~\cite{mercier_low-power_2022,kim2023, moosavifar2024}. This complexifies their integration with higher network communication layers, whereas standard-compatible WuRXs ease this process~\cite{roberts_268_2016,odelberg_fully_2024}. Second, WuRXs are often designed with passive circuitry to minimize power consumption. 
However, they are not able to achieve sensitivities below the \qty{-120}{dBm} of sensitivity of most LPWANs. Only a few designs~\cite{kim2021,kim2023} with custom modulations are capable of doing so with active circuitry.
Moreover, WuRXs have often been implemented separately from the main RX~\cite{mercier_low-power_2022,zhang2024}, as shown in Fig.~\ref{fig: intro}(a), or with additional off-chip components~\cite{moody2019,dissanayake_M108_Mems_2020}. Finally, even for the long-range design of~\cite{kim2023} that integrates the WuRX and the main RX in a single chip, the shared front-end circuit fixes the noise-power trade-off point of the entire receiver. This means that the WuRX performance will limit the performance of the main RX. These limitations are summarized in Fig.~\ref{fig: intro}(a).

In this paper, we propose a long-range programmable RF receiver \genRev{that exploits the sensitivity-power and sensitivity-latency trade-offs, as illustrated in Fig.~\ref{fig: intro}(b). This is enabled by programmable circuits in the RF front end that can dynamically modify their noise-power trade-offs to adapt to the wireless channel conditions.
When necessary, the average power consumption can be reduced at the expense of a reduction of the communication range~\cite{gonzalez2024}. Furthermore, \firstRev{the modulation parameters of LoRa, an LPWAN standard, are used to exploit the sensitivity-latency trade-off shown in Fig.~\ref{fig: intro}(b). LoRa uses chirps with different spreading factors~(SFs) to improve its noise robustness}. Additionally, a wake-up mode is integrated in the design. Developed for always-on receiver operation, it is an extreme design point of the noise-power trade-off. It is included in the receiver, with only a 4-\% silicon area overhead. The fabricated chip is able to properly receive LoRa packets sent by commercial chips with different SFs, making it the first LoRa-compatible receiver chip in the literature.}



The rest of the paper is organized as follows. Section~\ref{sec: SoA} gives an overview of the literature on wake-up and programmable receivers. Section~\ref{sec: Architecture} details the architecture of the receiver. Section~\ref{sec: Measurements} shows measurement results, and finally Section~\ref{sec: Conclusion} concludes the paper.

\section{Literature Review}
\label{sec: SoA}


In order to achieve sensitivity levels comparable to those specified by LPWAN standards, long-range WuRXs tend to adopt conventional heterodyne architectures~\cite{kim2021,zhang_-102dbm_2023,odelberg_fully_2024}, where the signal is downconverted from RF to either an intermediate frequency~(IF) or directly to baseband~(BB) through mixers\cite{mercier_low-power_2022}. This improves filtering and amplification capabilities and limits self-mixing issues~\cite{mercier_low-power_2022}. However, the power consumption increases significantly due to the required active circuitry, and the area overhead of WuRX cannot be neglected. 

Even though the WuRX and main RX architectures become increasingly similar, their sensitivities will differ. \firstRev{According to~\cite{semtech2020}, the demodulation of SF12 LoRa packets requires a minimum SNR of \qty{-20}{dB}. However, it is shown in~\cite{maistriaux2024} that their detection requires at least~\qty{-17}{dB} of SNR, which is \qty{3}{dB} higher. The demodulation and the detection of symbols are characterized with respect} to different metrics, i.e., the bit-error rate~(BER) for the main RX, and the MDR under a given false-alarm rate~(FAR) for the WuRX. In~\cite{kim2022}, a single front-end with a fixed NF is used for both the main RX and WuRX. Based on their analysis, the correlation performed on the WuS must be 32$\times$ longer than the data coding to obtain equivalent sensitivities. Indeed, longer WuS correlations improve the filtering of random noise prior to the wake-up detector. Therefore, sensitivity is improved at the expense of additional latency, as depicted in Fig.~\ref{fig: intro}(a). WuRXs most often employ low-cost 1-bit correlators~\cite{dissanayake2020,moody2019} that have poor efficiency for long WuSs~\cite{maistriaux2024}. More complex signal processing could improve the sensitivity by up to 11.8 dB~\cite{maistriaux2024}. However, multi-bit ADCs are required to limit quantization noise, which brings additional challenges since the WuRX power and area must be kept minimal. 

\genRev{An efficient WuRX design should simultaneously achieve a sufficient sensitivity while minimizing its power consumption}, as represented in Fig.~\ref{fig: intro}(a).
Due to the fundamental noise-power trade-off of analog front ends, the power consumption can be reduced if a degraded NF is accepted. This results in a sensitivity loss that can be compensated for by longer WuSs that increase the system latency. \genRev{The preamble structure of LoRa is already suitable for this purpose as it is composed of the same symbol repeated a configurable number of times. State-of-the-art preamble detection methods~\cite{tapparelLoRaPreambleDetection2023,kangLoRaModulationIoT2022} could thereby be leveraged.}

To minimize silicon area overhead, recent works include the wake-up capability directly in the main receiver. Moreover, the WuRX can be compatible with a standard, e.g., BLE~\cite{wang_dual-mode_2021} or NB-IoT~\cite{odelberg_fully_2024}, which significantly reduces the complexity of its integration at the physical and network layers. The short-range receiver in~\cite{roberts_268_2016} achieves this with a lower-power but higher-NF signal path in parallel to the main one with a limited sensitivity of \qty{-56.5}{dBm}. For long-range systems, the design in~\cite{kim2023} features a fixed-NF front end and a 4096-bit coded signal to reach sensitivities below \qty{-120}{dBm}. The use of a fixed-NF limits the available degrees of freedom that can be acted on between the wake-up and main-receiver modes because the power consumption in wake-up mode cannot be significantly reduced, nor can the data rate in main-receiver mode be increased.

Limitations of fixed-NF designs have also been pointed out for main receivers~\cite{gonzalez2024,sen2012,chatterjee2019,mukherjee2023}. RF receivers are conventionally designed with a fixed performance to guarantee a certain BER in the worst-case conditions of the wireless channel. However, these conditions are rarely encountered in practice because of the stochastic nature of the channel attenuation, which causes the RF receiver to overperform most of the time and thus to have a significant power overhead. It has been shown that programmable receivers can result in instantaneous power savings of a factor 2$\times$ compared to static designs~\cite{sen2012}, or in average power savings of 30 to \qty{75}{\%} in the case of wake-up receivers~\cite{mukherjee2023}. The design in~\cite{gonzalez2024} predicts \qty{46}{\%} of power savings on a complete LoRa receiver thanks to the dynamic configuration of the LNA at the expense of a 2-dB sensitivity degradation. With this goal, several programmable receivers or sub-blocks have been proposed, but they have either a power consumption in the tens of milliwatts that is prohibitive for IoT nodes using LPWANs \cite{sen2012,zhu2015,shafiee2022}, or a limited tuning range of only a few modes of operation \cite{zaini2016}. Recently, it has been shown that achieving both low power consumption and a wide range of configurability is possible in mixers \cite{das2025}.

\section{Proposed Programmable LoRa Receiver Architecture}
\label{sec: Architecture}

\genRev{In this work, we aim to tackle the limitations of previous receivers pointed out in the literature review. We propose the first standard-compatible long-range receiver with integrated wake-up mode. The high programmability in the main receiver mode allows it to adapt to the conditions of the wireless channel, which helps to reduce its average power consumption.}

\genRev{A wake-up mode intended for always-on operation is added with a specifically optimized power consumption. This avoids the aforementioned issues of having a fixed NF for the whole receiver. Both modes of operation share most of the hardware. This limits the area overhead linked with the addition of the wake-up mode.} The settling time between configuration change and mode was optimized to be negligible when compared to the period of a LoRa symbol, which results in seamless transition between wake-up mode and main receiving mode. \firstRev{This minimizes the symbol loss and thereby the impact of configuration change on the receiver performance.} 




We exploit the chirp spread-spectrum technique used in the LoRa modulation. Efficient FFT-based correlations are implemented in an ultra-low-power \firstRev{digital baseband~(DBB) processor} described in~\cite{maistriaux2024}. This allows the receiver to achieve the theoretical processing gain of LoRa with limited power overhead. \genRev{To save power, the NF is sacrificed in wake-up mode, which leads to a reduced sensitivity. To compensate for this, additional preamble detection methods~\cite{kang_lora_2023, kangLoRaModulationIoT2022} could easily be integrated to leverage the repeating symbol pattern of the LoRa preamble.}

\subsection{Architecture}

\begin{figure}[t]
\centering
\includegraphics[width=\columnwidth]{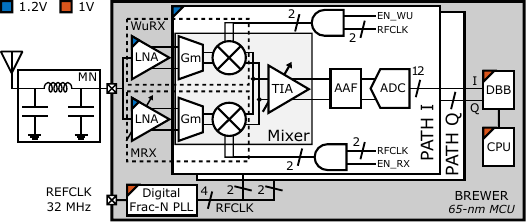}
\caption{Schematic of the proposed LoRa receiver.}
\label{fig: RXsch}
\end{figure}

The proposed direct-conversion RF receiver shown in \figurename~\ref{fig: RXsch} is implemented in the 65-nm technology from Taiwan Semiconductor Manufacturing Company~(TSMC) and supplied at \qty{1.2}{\volt}, except for the digital blocks, which are supplied at \qty{1}{\volt} to save switching power.
A direct-conversion architecture was chosen to simplify the receiver chain because simple low-pass filters can replace more complex image-rejection band-pass filters necessary in other architectures. The main disadvantage of the presence of flicker noise in the baseband is mitigated by the spread-spectrum modulation techniques used in LoRa. The despreading operation done at the receiver spreads the noise energy over the entire bandwidth \cite{torrieri2015}, limiting the effect of flicker noise on low frequencies.

The RF receiver includes an off-chip matching network for interfacing the antenna to a balun noise-canceling low-noise amplifier~(LNA). There are two LNAs, one for the wake-up mode and one programmable for the main-receiver mode, which can dynamically configure its noise-power trade-off. The LNA configuration for the wake-up mode could not be implemented with the other configurations due to the diverging specifications of the LNA for both modes of operation. The parallel paths generated at the LNAs converge inside the passive current-mode mixer. It is interfaced through a transconductance~(Gm) block and a transimpedance amplifier~(TIA). The latter offers another degree of programmability as it can also dynamically configure its noise-power trade-off. An all-digital fractional-N phase-locked loop~(ADPLL) generates the carrier frequency from a 32-MHz off-chip reference. The PLL is based on a quadrature ring oscillator, which generates the four phases of the carrier needed for the in-phase~(I) and quadrature~(Q) paths.
These paths consist of the passive current-mode mixer, an anti-aliasing filter~(AAF), and a 12-bit analog-to-digital converter~(ADC). Finally, synchronization and complete demodulation are performed in software, with a \firstRev{DBB that accelerates} the signal filtering, decimation, and demodulation of the LoRa signal.

Further details on the design of each block will be provided in the following sections.

\subsection{Programmable LNA}

\begin{figure}[t]
\centering
\includegraphics[width=\columnwidth]{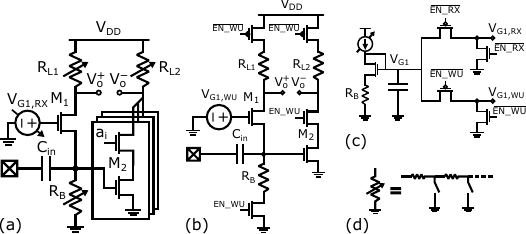}
\caption{(a) Schematic of the programmable LNA for the main-receiver mode. (b) Schematic of the LNA for the wake-up mode. (c) Gate voltage generation with a replica circuit. (d) Implementation of the variable resistor.}
\label{fig: LNAsch}
\end{figure}

\figurename~\ref{fig: LNAsch}(a) shows the schematic of the proposed programmable LNA for the main-receiver mode. Similar to \cite{gonzalez2024}, it is based on the noise-canceling balun LNA architecture~\cite{blaakmeer2008}. 
The common-gate amplifier on the left branch provides the gain on the positive path and sets the input impedance of the LNA, whereas the common-source branch on the right provides the gain on the negative path and is designed to cancel out the noise from the common-gate transistor M\textsubscript{1}. The combination of the balun and amplifier functions allows using fully differential blocks further in the chain to suppress common-mode noise while avoiding the insertion losses of a standalone balun needed to interface with the single-ended antenna.

Given that the common-gate transistor sets the input impedance of the LNA, a significant amount of power is usually spent on its biasing to match the 50-$\unit{\ohm}$ antenna impedance. Therefore, using a matching network to adapt the antenna impedance to a higher input impedance of the LNA allows for significant power savings. Moreover, the matching network adds passive gain to the chain, decreasing the impact of the noise figure~(NF) of the LNA on the global NF of the receiver, also saving power. A $\pi$-shaped matching network is used so parasitic capacitances due to chip packaging are in parallel with capacitor C\textsubscript{2} and thus limit the impact of their uncertainty on the S parameters of the LNA.

Four tuning knobs enable the dynamic change of the noise-power trade-off of the LNA in main-receiver mode. These tuning knobs are: the load resistors (R\textsubscript{L1} and R\textsubscript{L2}), the bias resistor R\textsubscript{B}, the gate voltage of transistor M\textsubscript{1}, V\textsubscript{G1}, and the width of transistor M\textsubscript{2}.
The gate voltage V\textsubscript{G1} is controlled by modifying the current flowing through a diode-connected replica of transistor M\textsubscript{1} with a simple current digital-to-analog converter~(DAC), as shown at the left of \figurename~\ref{fig: LNAsch}(c). The replica of resistor R\textsubscript{B} is up-sized to decrease the power consumed in this circuit. For tuning the bias and load resistances, series banks of resistors as shown in \figurename~\ref{fig: LNAsch}(d), are used, with the switches controlled by a one-hot-encoded digital word. In this case, a series bank of resistors is better than a parallel one because there is less parasitic capacitance at the output nodes of the LNA, reducing its impact on the gain. Finally, the width of transistor M\textsubscript{2} is controlled digitally through the signals a\textsubscript{i}, which activate parallel branches of common-source transistors of varying width.

For the wake-up mode, a stand-alone LNA was implemented as shown in \figurename~\ref{fig: RXsch}. One of the most important challenges of implementing different configurations in an LNA is the parasitic capacitance coming from the unused elements in the circuit. The challenge is particularly difficult when the extreme configurations have large differences in bias current. The low-power configurations, as the one needed for the wake-up mode, suffer from larger parasitic capacitance due to the large transistors needed for the high-power configurations, which heavily impacts the performance of the low-power ones. For this reason, a separate branch is required for the wake-up mode. \figurename~\ref{fig: LNAsch}(b) shows the schematic of the wake-up LNA. Four switches are used to enable this amplifier. When disabling either of the two LNAs, they consume less than \qty{25}{\nW}

As the block that generates the gate voltage of transistor M\textsubscript{1} occupies a significant area due to the decoupling capacitor, it is shared between the two LNAs, as shown in \figurename~\ref{fig: LNAsch}(c).


\begin{table}[t]
\centering
\caption{Ranges of the optimization parameters and performance constraints of the LNA.}
\label{tab: LNA}
\begin{adjustbox}{max width=\columnwidth}
\begin{tabular}{ll}
\begin{tabular}{cc}
\toprule
Parameter & Range \\
\midrule
M\textsubscript{1} gate voltage (V\textsubscript{G1})  & \num{0.58} to \qty{1.03}{\volt} \\
M\textsubscript{2} width & \num{8.5} to \qty{850}{\um} \\
R\textsubscript{L1} \& R\textsubscript{B} resistance & \num{5} to \qty{35}{\kilo$\Omega$} \\
R\textsubscript{L2} resistance & \num{50} to \qty{20000}{$\Omega$}\\
\bottomrule
\end{tabular}
&
\begin{tabular}{cc}
\toprule
Performance & Constraint \\
\midrule
S$_{11}$ & $\le$ \SI{-10}{dB} \\
Gain Imb. &  $\le$ \SI{2}{dB} \\
Phase Imb. &  $\le$ \SI{10}{\degree} \\
IIP3 &  $\ge$ \SI{-15}{dBm} \\
\bottomrule
\end{tabular}
\end{tabular}
\end{adjustbox}
\end{table}

Using a genetic optimization algorithm, as in \cite{gonzalez2024}, both LNAs were optimized to minimize the NF and the power consumption while maximizing the gain. The optimization was done under the constraints and parameter ranges summarized in Table~\ref{tab: LNA}. The matching network was also optimized with a genetic algorithm and fixed beforehand. The passive components were modeled through their S-parameters to account for their non-idealities in the optimization. The result for the optimum input impedance of the LNA was approximately \qty{1.5}{\kilo$\Omega$}, and thus the following components are needed to adapt it to the 50-$\unit{\ohm}$ antenna: an inductor of \qty{15.5}{\nH} and two capacitors of \qty{10}{\pF} and \qty{1.4}{\pF} for C\textsubscript{1} and C\textsubscript{2}, respectively.

\begin{figure}[t]
\centering
\includegraphics[width=\columnwidth]{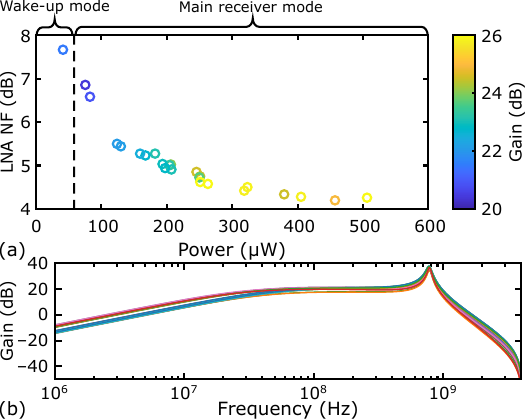}
\caption{\firstRev{Post-layout results of the LNAs. (a) Pareto front. (b) Frequency response for all configurations in the Pareto front.}}
\label{fig: LNApareto}
\end{figure}

The post-layout results of the Pareto front resulting from the genetic optimization algorithm are shown in \figurename~\ref{fig: LNApareto}\firstRev{(a)}. It is important to highlight that the configuration of the LNA can be changed dynamically to any of the points in the figure with a limited area overhead for its programmability. \firstRev{The frequency response of all the configurations is shown in \figurename~\ref{fig: LNApareto}(b).}
\firstRev{When considering the PVT variations, most performance metrics remain within $\pm$\qty{15}{\%} on average across the different configurations. However, the power consumption is the most impacted metric since the bias currents depend on the different resistances present in the circuit. The power can vary on average between \num{-80} and \qty{110}{\%}.}

In terms of start-up time, post-layout simulations show that the wake-up LNA needs \qty{74.48}{\us} to settle when enabled, while a change from the wake-up LNA to any configuration of the main-receiver LNA takes on average \qty{145}{\us}. This represents less than \qty{15}{\%} of a LoRa symbol using the highest data rate of LoRaWAN in Europe, the symbol period in this case is \qty{1.024}{ms} with an SF of 7 and a bandwidth of \qty{125}{kHz}.

\subsection{Programmable Mixer}

\begin{figure}[t]
\centering
\includegraphics[width=\columnwidth]{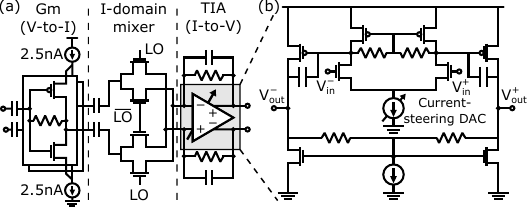}
\caption{Schematic of (a) the proposed programmable mixer, and (b) the fully differential amplifier used in the transimpedance amplifier of the mixer.}
\label{fig: MIXsch}
\end{figure}

The schematic of the mixer is shown in \figurename~\ref{fig: MIXsch}(a). A current-domain passive mixer architecture was chosen to minimize the flicker noise in the baseband in the direct-conversion receiver. AC coupling the switches makes their contribution to the flicker noise of the system negligible \cite{chehrazi2010}.
A transconductance~(Gm) block interfaces the LNA's voltage-domain output to the current-domain mixer. It consists of a current-starved resistive-feedback inverter. The starving allows the biasing of the inverter in weak inversion, reducing both its noise and its power consumption while increasing its gain.
Finally, the transimpedance amplifier~(TIA) converts the signal back to the voltage domain while providing a virtual ground for the switches and first-order filtering at \qty{500}{\kilo\hertz}.

\begin{figure}[t]
\centering
\includegraphics[width=\columnwidth]{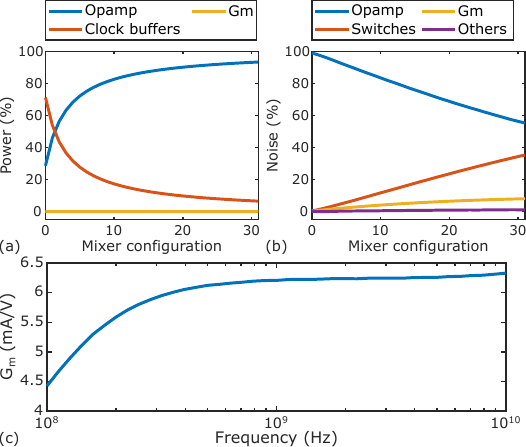}
\caption{(a) Power and (b) noise breakdown of the programmable mixer for all the inputs of the 5-bit DAC controlling the mixer configuration. \firstRev{(c) Frequency response of the Gm block.}}
\label{fig: MIXbreakdown}
\end{figure}

The Gm block and the clock buffers were designed such that the amplifier in the TIA dominates both the noise and the power consumption of the mixer, as shown in \figurename~\ref{fig: MIXbreakdown}. This makes it the perfect candidate for controlling the noise-power trade-off of the mixer. Furthermore, the optimization of the Gm and the clock buffers allowed their duplication for the wake-up branch at a negligible overhead to avoid additional switches in the RF path to change the operation mode of the receiver. Instead, this mode change is handled with AND gates, as shown in \figurename~\ref{fig: RXsch}. The TIA is shared between the two operation modes because it was designed with a large area to minimize its flicker noise. \firstRev{Even though the Gm block consumes very low power with the 2.5-nA bias current, \figurename~\ref{fig: MIXbreakdown}(c) shows that bandwidth is still sufficiently large for the needed frequency range.}

\begin{figure}[t]
\centering
\includegraphics[width=\columnwidth]{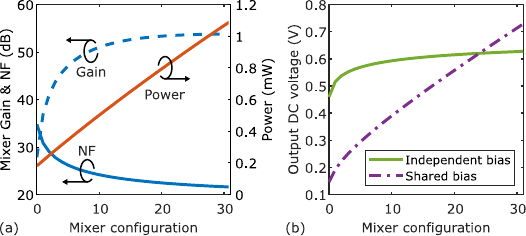}
\caption{(a) Post-layout results of the programmable mixer for all the inputs of the 5-bit DAC. (b) DC voltage of the output of the amplifier for shared or independent biasing of the two current sources.}
\label{fig: MIXresults}
\end{figure}

The schematic of the fully differential amplifier used inside the TIA is shown in \figurename~\ref{fig: MIXsch}(b) \cite{razavi2024}. It consists of a 2-stage Miller operational transconductance amplifier, each stage with a resistive common-mode feedback. Two current sources bias the amplifier: one biases the differential pair, and the other helps set the output DC voltage to half the supply. The bias current of the differential pair can be varied through a 5-bit current-steering DAC with a range between 2 and \qty{95}{\uA}. The resulting gain, NF, and power consumption of the mixer in post-layout simulations can be seen in \figurename~\ref{fig: MIXresults}(a). As for the LNA, the mixer configuration can be changed dynamically. 
The reason why the configuration is applied to the current source of the first stage is that varying the current source at the amplifier output by sharing the same biasing between both sources does not result in a good DC voltage at the output, as shown in \figurename~\ref{fig: MIXresults}(b). Fixing the current at the output by independently biasing the current sources results in a good DC voltage for all configurations with a negligible area overhead.

In wake-up mode, the mixer's lowest-power configuration is exclusively used to minimize power consumption. Post-layout simulations show that it takes \qty{39}{\us} to settle from power-on. Changing to main-receiver mode, which exploits all 32 configurations, takes between 45 and \qty{51}{\us}. Comparing again to a LoRa symbol in the maximum data rate, the mixer takes less than \qty{5}{\%} of a symbol to settle after any configuration change.

\subsection{All-Digital Phase-Locked Loop}
\label{sec: ADPLL}

Digital PLLs bring many benefits compared to their analog counterparts \cite{zhuang2014}.
First, technology scaling leads to a direct reduction of the area of the PLL control loop. The supply voltage can also be decreased to achieve a quadratic reduction of the power consumption. Second, a digital control provides more flexibility through the configuration of the filters, and it can be directly interfaced with the CPU. Lastly, whereas the voltage on the control node of a VCO might drift due to leakage currents, a disabled ADPLL keeps the control value of the digitally controlled oscillator~(DCO) unchanged as long as it is supplied, which enables operating in a low-power open-loop fashion where the frequency is maintained in the absence of temperature changes and voltage fluctuations. This can be exploited for avoiding a cold start.

\begin{figure*}[t]
\centering
\includegraphics[width=\textwidth]{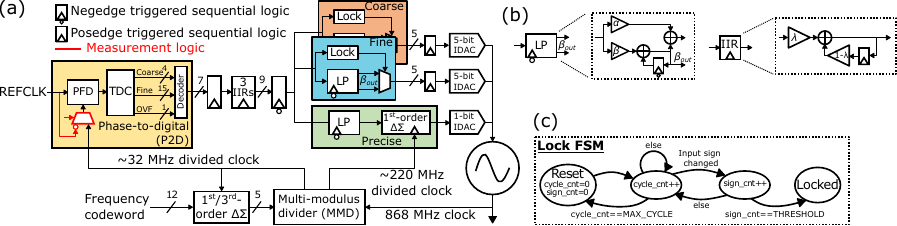 }%
\caption{(a) ADPLL architecture based on a current-starved ring oscillator. Three different current DACs controlled by distinct loop filters are used. (b)~Architecture of the LPs and IIR filters. (c)~FSM of the lock detection unit.}
\label{fig: ADPLL}
\end{figure*}

The selected architecture, depicted in \figurename~\ref{fig: ADPLL}(a), is referred to as a divider-based topology~\cite{zhuang2014}. It relies on a similar feedback loop as an analog charge-pump-based PLL. Analog blocks are replaced by their digital counterparts, with the charge pump translating into a time-to-digital converter~(TDC).
The architecture is similar to \cite{zhang2024}, but it is extended in several key aspects. First, the proposed PLL is fractional rather than integer, which controls the generated carrier frequency with higher accuracy. This requires the addition of a digital delta-sigma modulator~(DDSM). Second, the proposed ADPLL employs a current-starved ring oscillator rather than an LC DCO, which leads to a much more compact design at the cost of a higher intrinsic DCO phase noise.

As designing a DAC with a sufficient dynamic frequency range and resolution to maintain functionality under PVT variations while providing small enough frequency resolution is complex, ADPLLs generally rely on multiple parallel DACs~\cite{zhang2024,andersson2010,kwon2024}. This enables the design of simpler DACs that use only a few bits. A first one, with a coarser resolution, handles the adjustment to PVT variations, while one or multiple others provide a finer control. To achieve an even more precise control, a delta-sigma modulator can be added, increasing the effective resolution of the DCO~\cite{andersson2010}. This requires a low-pass filter to be added after the DDSM to prevent the high-frequency modulated noise from being directly injected into the DCO.

In LC DCOs, tunable capacitor banks are used for the digital control of the frequency. For ring oscillators, the design in \cite{kwon2024} relies on a cell-based DCO where the ring is decomposed into a configurable cell array. The oscillator frequency range and resolution are thereby tightly linked to its array, as more cells would be required to meet tighter specifications. Moreover, this topology prevents the use of a DDSM as no low-pass filter can be easily integrated.
\firstRev{Current-starved ring oscillators are attractive as the control current of the different current DACs can be easily combined. However, the dithered control of the DDSM needs to be filtered, which is not straightforward.}
 
To continue on the ADPLL architecture, the phase-to-digital converter~(P2D) performs the digitization of the delay between the edges of the reference clock and of the divided DCO clock in two steps. First, a phase-frequency detector~(PFD), combined with a multiplexer, translates this delay into a start-stop signal as presented in \cite{zhang2024}. The start signal launches the TDC operation while the stop signal samples its current value. The TDC architecture is based on a ring oscillator, similarly to~\cite{okazawa2017}. The TDC outputs a coarse value that corresponds to the number of completed ring oscillator cycles. This is achieved by connecting a counter to the oscillator's last stage. A fine result is obtained by sampling the state of the oscillator. A decoder then performs the thermometer-to-binary conversion of the fine output and adds it to the coarse output.

Three configurable infinite-impulse-response~(IIR) filters in series allow filtering the high-frequency modulated noise of the DDSM present at the TDC output. Their architecture, as well as that of the three parallel loop filters~(LPs), is depicted in \figurename~\ref{fig: ADPLL}(b). The selection and configuration of the weights of the LPs are based on the methodology presented in \cite{kratyuk2007}. The inclusion of the different blocks in the modeling of the phase noise, including the DDSM operating on the 1-bit current DAC, is based on the model developed in \cite{jiang2015}.

The ADPLL starts by activating the coarse loop control. Upon convergence of the DCO frequency, the control of the related DAC of the DCO is locked. The fine control loop is then activated, and the process repeats. Finally, the DDSM-controlled DAC is continuously active for precise phase control.  A lock detection unit identifies if the DCO frequency has settled by analyzing the TDC output. If its sign changes quickly, then the ADPLL has probably converged. This is translated in the FSM described in \figurename~\ref{fig: ADPLL}(c). A cycle counter is started to run for a configurable amount of time. When the TDC output sign changes, a second counter is incremented. If it goes over a threshold before the cycle counter expires, the control is locked. 

An additional advantage of having a digital architecture is that circuitry for post-silicon calibration can easily be added. In the case of the PLL, by measuring the actual gains of the TDC and the DCO, the LPs can be tuned to obtain the optimal performance. The circuitry for measuring the TDC gain is shown in red in \figurename~\ref{fig: ADPLL}(a). The PFD is fed with two opposed phases of the reference clock. The TDC then measures a known time difference equal to half the reference period, and its gain can be calculated by using the output values. For measuring the DCO gain with respect to its coarse and fine control words, the $\beta$\textsubscript{out} output of the LPs is obtained for two known frequency codewords.


\begin{figure}[t]
\centering
\includegraphics[width=\columnwidth]{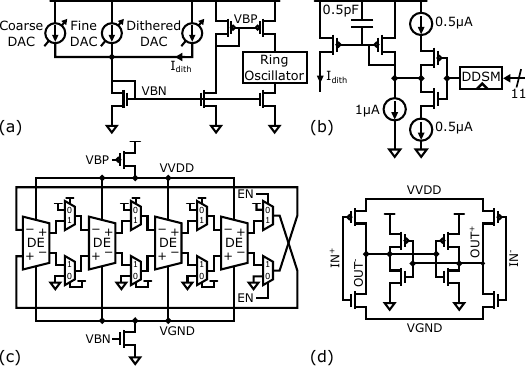}
\caption{Schematic of (a) the digitally controlled oscillator, (b) the sigma-delta current-domain dithered DAC, (c) the quadrature current-starved ring oscillator, and (d) the fully differential delay element~(DE).}
\label{fig: DCOsch}
\end{figure}

The DCO shown in \figurename~\ref{fig: DCOsch}(a) consists mainly of the quadrature current-starved ring oscillator~(CSRO) depicted in \figurename~\ref{fig: DCOsch}(c). A fully differential delay element (\figurename~\ref{fig: DCOsch}(d)) is used in the 4-stage ring oscillator to have access to the four phases needed in the I and Q differential mixers. As shown in~\cite{hofmann_PN_LoRa}, the LoRa modulation is robust against phase noise, especially its faster components. Ring oscillators can thus provide sufficiently low phase noise under the power budget of an IoT node~\cite{schramme2022} and do not require the area overhead of integrated inductors needed in LC tanks.

The CSRO is driven by three current-domain DACs: two 5-bit current-steering DACs and a sigma-delta current-domain DAC. Due to linearity requirements, a single DAC with the same range and resolution would be too complex. Instead, the complexity is transferred to the digital domain, as each DAC is controlled by a different digital filter. The coarse and fine DACs share the same architecture and implementation, but are biased with different currents to achieve a different range and precision.

\begin{figure}[t]
\centering
\includegraphics[width=\columnwidth]{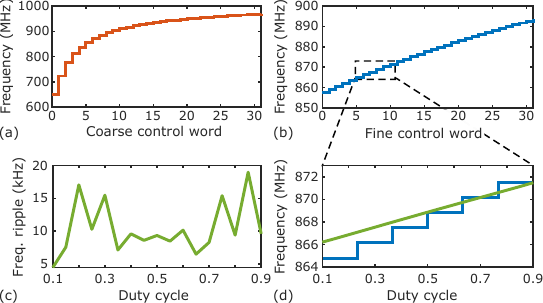}
\caption{Transfer function of the DCO for (a) the coarse DAC, with a fine control word of 0 and a duty cycle of 0.5, and (b) the fine DAC, with a coarse control word of 5 and a duty cycle of 0.5. \firstRev{(c) Peak-to-peak frequency ripple resulting from the sigma-delta DAC. (d) Transfer function of the DCO for the sigma-delta DAC, with a coarse control word of 5 and a fine control word of~8.}}
\label{fig: DCOTF}
\end{figure}

Each DAC serves a different purpose in the DCO. The coarse DAC compensates for the fabrication process and 5-\% voltage variations. The temperature variations are instead compensated with the current reference shared by the three DACs, which is proportional to absolute temperature, to reduce the necessary range of the coarse DAC.
The fine DAC is in charge of channel selection.
These two have a frequency resolution around their nominal operating point of 15 and \qty{1.3}{MHz}, respectively. Their complete transfer function can be seen in \figurename~\ref{fig: DCOTF}(a) and \ref{fig: DCOTF}(b).

However, the resolution of the fine DAC is insufficient to generate the exact needed frequency, causing spur tones in the phase noise profile when the fine control word toggles between different values to obtain the correct frequency on average.
Therefore, a DDSM is added to the loop to increase the resolution of the DCO. However, the same current-steering architecture of the other DACs cannot be used because the ripple caused by the dithered input would create spur tones in the phase noise profile again. 
Using capacitors to filter these ripples would affect the transient response of the other DACs.

Instead, we propose the \firstRev{new circuit} architecture shown in \figurename~\ref{fig: DCOsch}(b) to generate an average current depending on the 1-bit output signal of the DDSM while minimizing the resulting frequency ripple. 
This sigma-delta DAC dynamically changes the current passing through a correctly biased current mirror, which ensures the resulting average current is in the correct range. The dithering is done through a charge pump controlled by the duty cycle of the output signal of the DDSM. A capacitor connected between the gate and the source of the mirror transistors minimizes the frequency ripples. 
\figurename~\ref{fig: DCOTF}(c) and \ref{fig: DCOTF}(d) show the complete transfer function of this DAC and the resulting peak-to-peak frequency ripple, respectively.

\begin{figure}[t]
\centering
\includegraphics[width=\columnwidth]{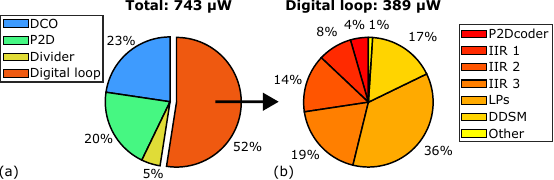}
\caption{(a) Power breakdown of the ADPLL. (b) Detailed breakdown of the digital control loop power consumption.}
\label{fig: ADPLL_PowerBreakdown}
\end{figure}

Except for the DCO, which was designed at the transistor level, and the DDSM and the digital filters, which were synthesized from HDL code, all other blocks in the ADPLL were designed using standard cells placed manually to guarantee an optimized layout in terms of symmetry and routing. \figurename~\ref{fig: ADPLL_PowerBreakdown}(a) shows the power breakdown from post-layout simulations and synthesis reports. The synthesized logic represents the greater contributor to the power consumption of the PLL, with the breakdown in terms of the different synthesized blocks shown in \figurename~\ref{fig: ADPLL_PowerBreakdown}(b). The different filters in the loop consume \qty{77}{\%} of the power of the synthesized blocks, even though their parameters were simplified to powers of 2 to avoid complex multiplication operations. 

\subsection{Analog Baseband Processing}
\label{sec: ABB}

\firstRev{Commercial LoRa modules have high blocking immunity, between 85 and \qty{100}{dB}~\cite{semtech2019abis,stmicroelectronics2022bis}. Achieving \qty{90}{dB} of attenuation at a 1-MHz offset from the carrier would need fifth-order filtering, i.e., a fourth-order AAF plus the first-order filtering done in the mixer. However, for power and area concerns, a second-order filter is preferred. Moreover, blockers can be further filtered in the digital domain, provided that they do not saturate the RF front end.} 

\firstRev{To implement the second-order filter, the Tow-Thomas biquadratic filter was shown for the AAF.}
The passive components and amplifiers in the AAF were designed such that the cut-off frequency will be greater than \qty{200}{kHz} even when considering mismatch and process, voltage, and temperature~(PVT) variations. This cut-off frequency is fixed sufficiently higher than the most common LoRa bandwidth, i.e., \qty{125}{kHz}, to avoid losing part of the signal in the presence of carrier frequency offsets between the transmitter and the receiver.

The ADC has a low-power successive-approximation architecture, as in \cite{favresse2025}. It features 12 bits of resolution and operates at a sampling frequency between \firstRev{\qty{125}{kS/s} and \qty{2}{MS/s} to enable up to 16$\times$ oversampling}. Thanks to self-synchronization, the ADC only needs the sampling clock. The charge-redistribution DAC uses fringing capacitors with an LSB unit of \qty{0.25}{fF}, resulting in a total input capacitance of only \qty{1.2}{pF} per side, including parasitics. This saves DAC switching power and driver loading. A conventional DAC switching scheme is applied, as it only needs the supply voltage as a voltage reference. Each ADC consumes between \qty{671.3}{nW} and \qty{10.3}{\uW}, depending on the sampling frequency.

\subsection{Digital Baseband Processing and Software}
\label{sec: DBB}

A DBB processor performs the signal filtering, decimation, and finally the LoRa symbol demodulation, which is the most computationally intensive operation underlying the LoRa physical layer. Indeed, LoRa can operate at SNRs down to \qty{-20}{dB} by correlating incoming symbols and reference chirps that are up to 4096-point long. A programmable FIR filter presented in \cite{xhonneux2021a} further reduces the in-band noise and decimates the signal from \qty{2}{MHz} to \qty{125}{kHz}. Next, the chirp-spread spectrum modulation underlying the LoRa standard requires dechirping the incoming signal by multiplying it with the reference chirp. To avoid storing all possible reference chirps, a modified version of the lightweight chirp generator of~\cite{nguyen2019} is used. The result is then stored in two ping-pong buffers before performing the fast Fourier transform~(FFT) as described in. The ping-pong buffer guarantees the continuous operation of the system. The FFT architecture is based on the design of \cite{maistriaux2024}. However, a higher radix-2\textsuperscript{3} processing unit described in~\cite{maistriaux2026} is used to reduce the FFT execution time and energy consumption. Finally, the DBB processor identifies the index and value of the maximum magnitude among all FFT bins. This information can then be accessed by the Cortex-M4 CPU integrated on the chip to carry out the next step of the demodulation.

With most intensive operations carried out by the DBB processor, the Cortex-M4 only handles the synchronization and packet decoding that can be easily performed in software. The synchronization process from \cite{xhonneux2021} is used to \firstRev{correct both the sample timing offset~(STO) and the carrier frequency offset~(CFO). It relies on the maximum of the FFT outputs and the adjacent values. This synchronization process is facilitated by the DBB since it has already identified the location of the maximum. The STO correction is performed by changing the decimation index of the FIR filter that operates on the $16\times$ oversampled signal.} The decoding of the LoRa packet studied in \cite{xu2022b} is performed after the demodulation of all transmitted symbols. \firstRev{The complete LoRa physical layer is thereby implemented in the proposed receiver. The addition of LoRaWAN as the medium access control~(MAC) layer could be made fairly easily in software.}

\genRev{The wake-up mode performs the same dechirping-FFT process. The ADC oversampling and the FIR could be deactivated but this does not result in any significant power savings. Both were thus kept for additional noise filtering. The maximum magnitude of the FFT outputs is compared to a predefined threshold to declare a detection. The threshold is set up according to~\cite{maistriaux2024}.}

\genRev{Currently, a single LoRa symbol is used to perform the detection, which limits the sensitivity of the wake-up mode. Multiple methods could be used to leverage the repetition of upchirps in the preamble, whose number is configurable in LoRa commercial chips~\cite{semtech2020}. In~\cite{kangLoRaModulationIoT2022}, the input samples are averaged, which is equivalent to averaging the complex FFT output results. In theory, it provides a processing gain proportional to the number of symbols used, e.g., \qty{12.04}{dB} for 16 symbols used. However, this method is very sensitive to the unknown fractional CFO~\cite{tapparelLoRaPreambleDetection2023}. In~\cite{kang_lora_2023}, a detection is assessed if $M$ symbols over $K$ lead to individual detections. This method is less effective, with a processing gain only of \qty{5.90}{dB} for 16 symbols used based on our estimations.}

\genRev{Such signal processing methods could easily be integrated in the proposed receiver. Indeed, it would require no modification to the analog front end. Moreover, the computationally intensive tasks of the process, i.e., the FIR, the dechirping, and the FFT, would require no modification. The additional logic would only operate on the on the output data that has a much lower datarate. The impact on the power consumption should thus be very limited. }

\section{Measurement results}
\label{sec: Measurements}

\begin{figure}[t]
\centering
\includegraphics[width=\columnwidth]{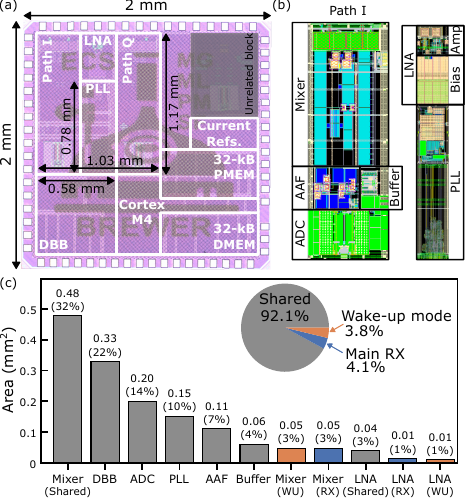}
\caption{(a) Chip microphotograph. (b) Layout of the RF front end. \firstRev{(c) Area breakdown.}}
\label{fig: Chip}
\end{figure}

The proposed programmable LoRa receiver occupies an area of \qty{1.6}{mm\squared} on the 2$\times$\qty{2}{mm\squared} Cortex-M4 microcontroller~(MCU) chip codenamed BREWER, as shown in \figurename~\ref{fig: Chip}(a).
The MCU also contains two SRAMs, a 32-kB program memory~(PMEM) and a 32-kB data memory~(DMEM), as well as current references for biasing the analog blocks.
\figurename~\ref{fig: Chip}(b) shows the layout of the RF front end. Only the path I is shown as the path Q is identical.

In terms of the area overhead of including the wake-up \firstRev{mode} in the LoRa receiver, it mainly impacts the LNA. The stand-alone wake-up LNA corresponds to \qty{10.8}{\%} of the \qty{0.0558}{mm\squared} LNA block shown in \figurename~\ref{fig: Chip}(a). The area impact of the duplicated blocks in the mixer is negligible, as the TIA largely dominates the area to reduce its flicker noise. \firstRev{\figurename~\ref{fig: Chip}(c) shows that the duplicated blocks for the wake-up mode account only for \qty{3.8}{\%} of the total area of the receiver.}

\begin{figure}[t]
\centering
\includegraphics[width=\columnwidth]{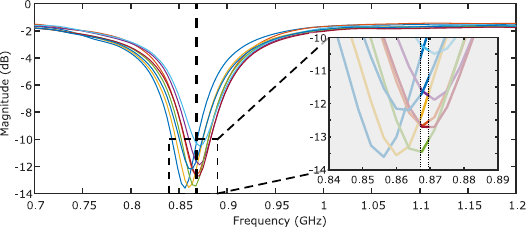}
\caption{Measured S\textsubscript{11} parameter on 8 different chips. The inset shows a zoom and highlights the frequency range used by LoRa.}
\label{fig: MeasS11}
\end{figure}

\figurename~\ref{fig: MeasS11} shows the measured S\textsubscript{11} across eight chips, showing only small variability in the peak's frequency. The inset in the figure shows that, inside the frequency band of LoRaWAN between 867.1 and \qty{868.8}{MHz}, all chips have an S\textsubscript{11} below \qty{-10}{dB}, which is the constraint used when designing the LNA. The different configurations of the LNA have a mean S\textsubscript{11} of \qty{-11.8}{dB} with a relative standard deviation of \qty{1.2}{\%}.

\begin{figure}[t]
\centering
\includegraphics[width=\columnwidth]{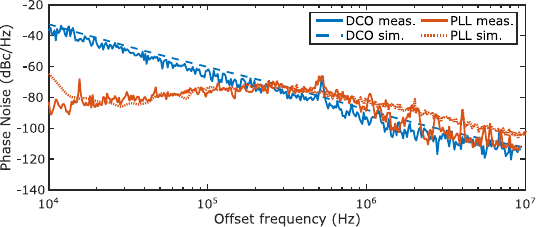}
\caption{Measured phase noise for the free-running DCO and the ADPLL. Dashed lines show the simulation results for comparison.}
\label{fig: PhNmeasured}
\end{figure}

The measured phase noise profiles of the free-running DCO and the ADPLL are shown in \figurename~\ref{fig: PhNmeasured} compared to the simulated profiles. There is a good agreement between the simulated and measured phase noise for both, showcasing the advantage of the digital PLL, in which we can measure the actual gains and adjust the loop parameters accordingly.

\begin{figure}[t]
\centering
\includegraphics[width=\columnwidth]{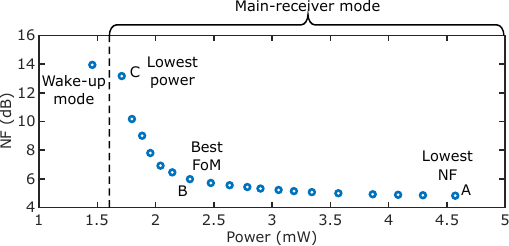}
\caption{Pareto front of the LoRa RF receiver \firstRev{using measured power consumption and simulated NF}. Configurations A, B, and C are highlighted.}
\label{fig: ParetoRFFE}
\end{figure}

\firstRev{When sending LoRa packets with SF 6 to 12 and bandwidths of 62.5 and \qty{125}{kHz}, the fabricated receiver correctly receives them. This validates the LoRa-compatibility of the design. However, this required an input power above \qty{-90}{dBm}, significantly higher than the expected sensitivity}. This is due to a random offset voltage that appears at the mixer’s output, which, unfortunately, was DC-coupled to the AAF input in the prototyped chip. As the gain of the AAF is significant, the amplified DC mixer offset brings the AAF output voltage close to saturation. To avoid this NF degradation, the DC coupling between the mixer and the AAF should be replaced by AC coupling\firstRev{, or a DC offset cancellation loop should be implemented.} 
\firstRev{As the DC offset at the mixer output causes the AAF to saturate, both the gain and the noise performance of the receiver chain are affected, preventing a reliable measurement of the NF or the sensitivity.}

Assuming that we can reproduce the noise performance we observe in post-layout simulations by fixing the problems linked to the offset voltage, we would obtain the Pareto front shown in \figurename~\ref{fig: ParetoRFFE} for the complete receiver. The power consumption corresponds to the measured one to have a fairer comparison to the state of the art. The programmable LNA and mixer allow a range of NF between 4.57 and \qty{13.17}{dB} for a corresponding power consumption between 4.57 and \qty{1.71}{mW}. This means that the receiver is able to save \qty{62.5}{\%} of power consumption by degrading its sensitivity by \qty{8.3}{dB}. \firstRev{In wake-up mode, the receiver can go down to \qty{1.49}{mW} for an NF of \qty{13.95}{dB}.}

\begin{figure}[t]
\centering
\includegraphics[width=\columnwidth]{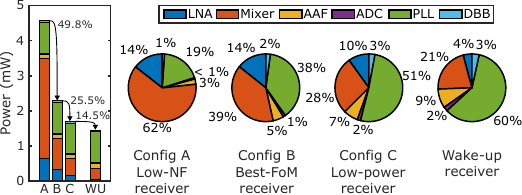}
\caption{\firstRev{Breakdown of the measured power consumption} for the wake-up mode and the three highlighted configurations of the LoRa receiver. The percentage corresponding to the LNA, mixer, AAF, and ADC is based on post-layout simulations.}
\label{fig: PowerBreakdown}
\end{figure}

The power breakdown of three configurations is given in \figurename~\ref{fig: PowerBreakdown}, corresponding to the two extreme configurations and one that approaches the noise performance of the commercial SX1276 chip from Semtech, labeled as configurations A, B, and C for easy reference. The power breakdown reveals that the PLL and mixers clearly dominate the receiver's power consumption, followed by the LNA. This reinforces the idea that exploiting the noise-power trade-off of these blocks should be prioritized in programmable receivers. The power breakdown in wake-up mode shows that the PLL is the dominant block. \firstRev{As already identified in Fig.~\ref{fig: ADPLL}, this is partly due to the four cascaded filters in the control loop}. The PLL should thus be the focus of optimization to reduce the power consumption in wake-up mode below \qty{1}{mW}. This could be achieved by scaling to a more advanced technology node, by modifying the ADPLL to include a specific low-power configuration for the WuRX, or by designing a simple charge-pump PLL optimized for low power consumption. Additionally, it was shown in~\cite{hofmann_PN_LoRa} that LoRa is mainly sensitive to slow phase noise components. Innovative techniques, such as injection locking~\cite{li_injection_lock}, could be studied to use free-running oscillators with bounded low-frequency phase noise. Duty-cycling techniques, largely used in state-of-the-art WuRXs, could also be considered to reduce the power consumption in wake-up mode. This could be enabled by the small settling times of the different blocks in the chain. 

%

\begin{figure}[t]
\centering
\includegraphics[width=\columnwidth]{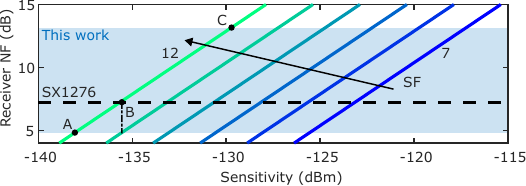}
\caption{NF required at the receiver to attain the sensitivity of LoRa as a function of the SF. The dotted line represents the mean NF of the SX1276 chip from Semtech~\cite{semtech2020}. Two receiver configurations are highlighted, exhibiting the best and worst NF, respectively.}
\label{fig: NFvsSensitivity}
\end{figure}

The sensitivity achieved for a given receiver NF is given by the following expression
\begin{equation}
    \text{Sensitivity [dBm]} = 10\log(kTB) + NF + SNR_{\text{min}},
\end{equation}
where $k$ is Boltzmann's constant, $T$ is the absolute temperature in Kelvin, and $B$ is the bandwidth of the signal of interest. The $SNR_{\text{min}}$ represents the minimum SNR needed to demodulate a LoRa symbol. The SF directly affects the $SNR_{\text{min}}$ because \qty{2.5}{dB} more are needed each time the SF is decreased by one~\cite{semtech2020}, \firstRev{with a $SNR_{\text{min}}$ of \qty{-7.5}{dB} in SF7}.
The resulting curves of the required NF versus sensitivity for a 125-kHz bandwidth are shown in \figurename~\ref{fig: NFvsSensitivity}. The dotted line indicates the mean NF for all SFs of the SX1276 chip~\cite{semtech2020} at this bandwidth. The proposed receiver can be changed dynamically in a range from point A to point C, saving more than 2.5$\times$ power. 
However, as the SF can be reduced when the SNR improves by \qty{2.5}{dB}, resulting in a shorter packet, more power can be saved at the network level by reducing the packet energy at the transmitter than by further degrading the NF at the receiver. Therefore, the configurations that will be used more frequently are limited to configuration B, which saves \qty{49.8}{\%} of power consumption compared to configuration A. In the particular case of SF7, where a lower SF is not available in the LoRaWAN protocol to further gain in transmitter energy, the receiver can still save \qty{25.5}{\%} more power by going up to configuration~C when the observed SNR is at its best.

\begin{table*}[!t]
\centering
\begin{adjustbox}{max width=\textwidth}
\begin{threeparttable}
\renewcommand{\arraystretch}{1.05}
\caption{Comparison to state-of-the-art low-power receivers.}
\label{tab: RX_SoTA}
\centering
\begin{tabular}{l *{4}{c @{\hspace{0.75\tabcolsep}}} c *{2}{c @{\hspace{0.75\tabcolsep}}} c *{2}{c @{\hspace{0.75\tabcolsep}} c}  *{4}{c} @{\hspace{0.75\tabcolsep}} c}
\toprule
\multicolumn{1}{l}{} & \multicolumn{14}{c}{Long Range} & \multicolumn{3}{c}{Short Range} \\
\cmidrule(lr){2-15}
\cmidrule(lr){16-18}
 & \multicolumn{5}{c}{\bfseries This work} & 
\cite{semtech2019abis} & \cite{semtech2013bis} & \cite{stmicroelectronics2022bis} &
\multicolumn{2}{c}{\cite{kim2022}} & \multicolumn{2}{c}{\cite{kim2021}} & \cite{raja2020} & \cite{lachartre2017} &
\cite{scolari2024} & \multicolumn{2}{c}{\cite{masnadishirazi2021}} \\
Publication & \multicolumn{5}{c}{} & \multicolumn{3}{c}{\multirow{2}{*}{Datasheet}} &
\multicolumn{2}{c}{ISSCC} & \multicolumn{2}{c}{JSSC} & SSC-L & ISSCC &
ISSCC &  \multicolumn{2}{c}{TMTT} \\
Year & \multicolumn{5}{c}{2025} & & & &
\multicolumn{2}{c}{2022} & \multicolumn{2}{c}{2021} & 2017 & 2017 &
2024 &  \multicolumn{2}{c}{2021} \\
 \midrule
\multirow{2}{*}{Technology} & \multicolumn{5}{c}{\SI{65}{\nm}} & \multirow{2}{*}{N/A} & \multirow{2}{*}{N/A} & \multirow{2}{*}{N/A} &
\multicolumn{2}{c}{\SI{55}{\nm}} & \multicolumn{2}{c}{\SI{55}{\nm}} & \SI{0.13}{\um} & \SI{65}{\nm} &
\SI{22}{\nm} & \multicolumn{2}{c}{\SI{40}{\nm}} \\
 & \multicolumn{5}{c}{bulk} & & & & 
 \multicolumn{2}{c}{bulk} & \multicolumn{2}{c}{bulk} & bulk & bulk &
 FD-SOI & \multicolumn{2}{c}{bulk} \\
Standard & \multicolumn{5}{c}{LoRa} & LoRa & -- & LoRa &
\multicolumn{2}{c}{--} & \multicolumn{2}{c}{--} & -- & Sigfox &
BLE & \multicolumn{2}{c}{BLE} \\
Modulation & \multicolumn{5}{c}{CSS} & CSS & FSK & CSS &
\multicolumn{2}{c}{BFSK} & \multicolumn{2}{c}{SC-BFSK} & (G)FSK & DBPSK &
GFSK & \multicolumn{2}{c}{GFSK} \\
Freq. [GHz] & \multicolumn{5}{c}{0.868} & 0.868 & 0.869 & 0.868 & 
\multicolumn{2}{c}{0.924} & \multicolumn{2}{c}{0.9} & 0.42-0.47 & 0.85-0.92 &
2.4 & \multicolumn{2}{c}{2.4} \\
Supply [V] & \multicolumn{5}{c}{1.16/1.2} & 3.3 & 3.3 & 3.3 & 
\multicolumn{2}{c}{0.94/1.3} & \multicolumn{2}{c}{0.9/1} & 1.2/1.5 & 3.3 &
0.65 & \multicolumn{2}{c}{1} \\
Power [mW] & \multicolumn{2}{c}{1.71} & 2.29 & \multicolumn{2}{c}{4.57} & 35.64 & 9.9 & 15.91 &
 \multicolumn{2}{c}{0.78} & 0.88 & 19.88 & 13.5 & 14.5 &
 2.95 & 0.98 & 3.48 \\
NF [dB] & \multicolumn{2}{c}{13.17\tnote{*}} & 5.98\tnote{*} & \multicolumn{2}{c}{4.83\tnote{*}} & 5.89\tnote{$\dagger$} & 7 & 4.89\tnote{$\dagger$} & \multicolumn{2}{c}{6} & \multicolumn{2}{c}{6} & 3.5 & N/A &
 N/A & \multicolumn{2}{c}{5.2} \\ 
DR [kbps] & 0.37 & 6.8 & 0.37 & 0.37 & 6.8 & 0.37 & 25 & 0.37 &
31.3 & 250 & 50 & 0.39 & 4.8 & 0.1 &
1000 & \multicolumn{2}{c}{1000} \\
Sens. [dBm] & \si{-130} & \si{-117} & \si{-137} & \si{-138} & \si{-126} & \si{-137} & \si{-107} & \si{-138} & 
\si{-107} & \si{-100} & \si{-106} & \si{-123} & \si{-121} & \si{-136}  &
\si{-98} & \multicolumn{2}{c}{\si{-95}} \\
RX FoM\tnote{$\ddagger$} & 183 & 183.2 & 188.9 & 187.1 & 187.3 & 177.1 & 171 & 181.6 &
183 & 185 & 183.5 & 165.9 & 176.5 & 174.4 & 
183.5 & 185.2 & 179.7 \\
Area \firstRev{[mm\textsuperscript{2}]} & \multicolumn{5}{c}{1.6} & N/A & N/A & N/A &
\multicolumn{2}{c}{3.75} & \multicolumn{2}{c}{5} & 10.24 & 3.76 &
1.03 & \multicolumn{2}{c}{0.7} \\
\bottomrule
\end{tabular}
\begin{tablenotes}
\item[*] Post-layout simulations.
\item[$\dagger$] Calculated from datasheet.
\item[$\ddagger$] RX FoM = $-\text{Sensitivity} - 10\log(\text{Power}) + 10\log(\text{Data rate})$.
\end{tablenotes}
\end{threeparttable}
\end{adjustbox}
\end{table*}

\begin{figure}[t]
\centering
\includegraphics[width=\columnwidth]{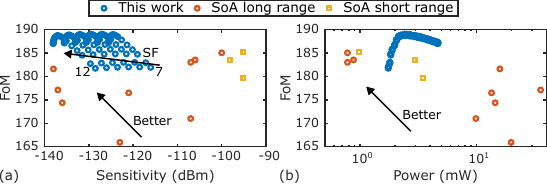}
\caption{Comparison to the state of the art in terms of the receiver FoM as a function of (a) sensitivity, and (b) power consumption.}
\label{fig: FoM_RX}
\end{figure}

Table~\ref{tab: RX_SoTA} shows the comparison of the proposed LoRa receiver with state-of-the-art long- and short-range low-power radio receivers.
For a quantitative comparison, we use the following figure of merit~(FoM):
\begin{equation}
\text{RX FoM} = -\text{Sensitivity} - 10\log(\text{Power}) + 10\log(\text{DR}),
\end{equation}
where DR is the data rate.
The resulting FoM can be visualized as a function of the sensitivity and the power consumption in \figurename~\ref{fig: FoM_RX}. The proposed programmable receiver achieves the best FoM compared to all other designs. Compared to long-range receivers with a sensitivity below \qty{-110}{dBm}, it also achieves the lowest power consumption. Receivers consuming less than \qty{1}{mW} have a good FoM, but their sensitivity is limited to \qty{-107}{dBm}, which is not compatible with LoRa. Furthermore, compared to commercial LoRa receivers, the proposed design consumes 3.5 to 8$\times$ less power for a similar sensitivity.
\firstRev{However, it must be highlighted that this FoM is very selective in terms of the performance metrics it takes into consideration for the comparison, even though it is commonly used in the literature. For instance, the receiver's linearity, as measured by the input third-order intercept point (IIP3) metric, is absent from this FoM, even though it is key to good blocking immunity. We decided to leave aside the IIP3 optimization in the proposed prototype to simplify the design. This could partly explain the great power savings compared to commercial LoRa chips for similar sensitivities. Future works on programmable receivers should focus on optimizing the IIP3 to add robustness to interference, possibly trying to exploit a linearity-power trade-off as was done for the noise in this work.}

\begin{table*}[!t]
\centering
\begin{adjustbox}{max width=\textwidth}
\begin{threeparttable}
\renewcommand{\arraystretch}{1.05}
\caption{Comparison to state-of-the-art long-range wake-up receivers.}
\label{tab: WU_SoTA}
\centering

\begin{tabular}{l *{2}{c @{\hspace{0.75\tabcolsep}} c} c c *{2}{c @{\hspace{0.75\tabcolsep}} c} *{3}{c @{\hspace{0.75\tabcolsep}}}}

\toprule
 & \multicolumn{2}{c}{\bfseries This work} & \multicolumn{2}{c}{\cite{kim2023}} & \cite{odelberg_fully_2024} & \cite{moosavifar2024} & \multicolumn{2}{c}{\cite{dissanayake2020}} & \multicolumn{2}{c}{\cite{zhang2024}} & \multicolumn{3}{c}{\cite{dissanayake2022}} \\
Publication & \multicolumn{2}{c}{} & \multicolumn{2}{c}{JSSC} & ISCAS & JSSC & \multicolumn{2}{c}{VLSI} & \multicolumn{2}{c}{OJSSCS} & \multicolumn{3}{c}{JSSC} \\
Year & \multicolumn{2}{c}{2025} & \multicolumn{2}{c}{2023} & 2024 & 2024 & \multicolumn{2}{c}{2020} & \multicolumn{2}{c}{2024} & \multicolumn{3}{c}{2022} \\
 \midrule
 
Standard & \multicolumn{2}{c}{LoRa} & \multicolumn{2}{c}{N/A} & NB-IoT & N/A & \multicolumn{2}{c}{N/A} & \multicolumn{2}{c}{N/A} & \multicolumn{3}{c}{N/A} \\

Modulation & \multicolumn{2}{c}{CSS} & \multicolumn{2}{c}{BFSK} & OFDM & Dual-chirp OOK & \multicolumn{2}{c}{CE-OOK} & \multicolumn{2}{c}{BFSK} & \multicolumn{3}{c}{CE-OOK} \\

Frequency & \multicolumn{2}{c}{\qty{868}{MHz}} & \multicolumn{2}{c}{\qty{900}{MHz}} & \qty{868}{MHz} & \qty{433}{MHz} & \multicolumn{2}{c}{\qty{434}{MHz}} & \multicolumn{2}{c}{\qty{2.4}{GHz}} & \multicolumn{3}{c}{\qty{2.4}{GHz}} \\

Bandwidth & \multicolumn{2}{c}{\qty{125}{kHz}} & \multicolumn{2}{c}{\qty{62.5}{kHz}} & \qty{180}{kHz} & \qty{440}{kHz} & \qty{200}{Hz} & \qty{20}{Hz} & \multicolumn{2}{c}{\qty{30}{kHz}} & \multicolumn{3}{c}{\qty{16.4}{kHz}} \\

Data Rate & \multicolumn{2}{c}{\qty{125}{kS/s}} & \multicolumn{2}{c}{\qty{31.25}{kbps}} & N/A & \qty{2.5}{kbps} & \qty{100}{bps} & \qty{10}{bps} & \multicolumn{2}{c}{\qty{15.4}{kbps}} & \multicolumn{3}{c}{\qty{8.192}{kbps}} \\

Architecture & \multicolumn{2}{c}{Direct Conversion} & \multicolumn{2}{c}{Direct Conversion} & Heterodyne & Self-mixing LNA & \multicolumn{2}{c}{LNA-ED-first} & \multicolumn{2}{c}{Heterodyne} & \multicolumn{3}{c}{Uncertain IF} \\

\multirow{2}{*}{Technology} & \multicolumn{2}{c}{\qty{65}{nm}} & \multicolumn{2}{c}{\qty{55}{nm}} & \multirow{2}{*}{\qty{28}{nm}}& \qty{28}{nm} & \multicolumn{2}{c}{\qty{65}{nm}} & \multicolumn{2}{c}{\qty{65}{nm}} & \multicolumn{3}{c}{\qty{65}{nm}} \\
 & \multicolumn{2}{c}{bulk} & \multicolumn{2}{c}{bulk} &  & bulk & \multicolumn{2}{c}{bulk} & \multicolumn{2}{c}{bulk} & \multicolumn{3}{c}{bulk} \\
 

Off-chip & \multicolumn{2}{c}{32-MHz XO} & \multicolumn{2}{c}{\multirow{2}{*}{Inductor + RC filter}} & \multirow{2}{*}{12-MHz XO} & \multirow{2}{*}{4 inductors} & \multicolumn{2}{c}{\multirow{2}{*}{MN}} & \multicolumn{2}{c}{32-kHz XO} & \multicolumn{3}{c}{32-kHz XO} \\
components\tnote{$\dagger$} & \multicolumn{2}{c}{+ MN} &  & & & & & & \multicolumn{2}{c}{+2-MHz XO} & \multicolumn{3}{c}{+1-MHz XO} \\
 
Supply (V) & \multicolumn{2}{c}{1.12/1.2} & \multicolumn{2}{c}{0.94/1.3} & 0.9 & 0.5/0.8 & \multicolumn{2}{c}{N/A} & \multicolumn{2}{c}{0.6/0.75/0.88/1} & \multicolumn{3}{c}{0.6/1} \\
 
Area (\unit{mm\squared}) & \multicolumn{2}{c}{1.6} & \multicolumn{2}{c}{3.75} & 1.08 & 0.75 & \multicolumn{2}{c}{4.5} & \multicolumn{2}{c}{5.22} & \multicolumn{3}{c}{2.24} \\

WU code length & \qty{128}{pts} & \qty{4096}{pts} & \multicolumn{2}{c}{\qty{4096}{pts}} & \qty{132}{bits} & \qty{24}{bits} & \multicolumn{2}{c}{\qty{26}{bits}} & \multicolumn{2}{c}{\qty{64}{bits}} & \multicolumn{3}{c}{\qty{32}{bits}} \\

WU code duration (ms) & 1 & 32 & \multicolumn{2}{c}{131} & 0.733 & 9.6 & 260 & 2600 & \multicolumn{2}{c}{4} & \multicolumn{3}{c}{3.91} \\

WuS repetition & \multicolumn{2}{c}{No} & \multicolumn{2}{c}{15$\times$} & No & 1$\times$ & \multicolumn{2}{c}{No} & 1$\times$ & 4000$\times$ & 1$\times$ & N/A & N/A \\

Latency & \qty{1}{ms} & \qty{32}{ms} & \qty{130}{ms} & \qty{1.97}{s} & \qty{785}{\us} & \qty{9.6}{ms} & \qty{260}{ms} & \qty{2.6}{s} & \qty{4}{ms} & \qty{16}{s} & \qty{3.9}{ms} & \qty{10}{ms} & \qty{1}{s} \\

Power (\unit{\uW}) & 1458 & 1508 & 781 & 114 & 2400 & 110 & 2.17 & 0.26 & 1739 & 2.2 & 539 & 20.9 & 0.9 \\

Sensitivity (dBm) & \num{-111}\tnote{*} & \num{-126}\tnote{*} & \multicolumn{2}{c}{\num{-124}} & \num{-107} & N/A & \multicolumn{2}{c}{\num{-99}} & \multicolumn{2}{c}{\num{-102}} & \multicolumn{3}{c}{\num{-91.5}} \\

WU FoM (dB)\tnote{$\ddagger$} & 169.8 & 169.2 & 163.9 & 160.5 & 164.2 & N/A & 161.5 & 160.7 & 153.6 & 146.5 & 150.3 & 156.4 & 149.1 \\

Duty cycling & \multicolumn{2}{c}{AO} & AO & PLDC & AO & AO & \multicolumn{2}{c}{BLDC} & AO & PLDC & AO & WPDC & WPDC \\
 
\bottomrule
\end{tabular}
\begin{tablenotes}
\textbf{AO}:Always On \textbf{BLDC}: Bit-Level Duty-Cycling \textbf{CE}:Channel-Embedded \textbf{ED}: Envelope Detector \textbf{IF}: Intermediate Frequency \textbf{MN}: Matching Network \textbf{OFDM}: Orthogonal Frequency-Division Multiplexing \textbf{OOK}: On-Off Keying \textbf{PLDC}: Packet-Level Duty-Cycling \textbf{WPDC}:Within-Packet Duty-Cycling \textbf{XO}: Crystal Oscillator
\item[*] Post-layout simulations.
\item[$\dagger$] XOs are reported if their frequency is higher than \qty{1}{MHz}.
\item[$\ddagger$] WU FoM = $-\text{Sensitivity} - 10\log(\text{Power}) - 10\log(\text{Latency})$.
\end{tablenotes}
\end{threeparttable}
\end{adjustbox}
\end{table*}

\begin{figure}[t]
\centering
\includegraphics[width=\columnwidth]{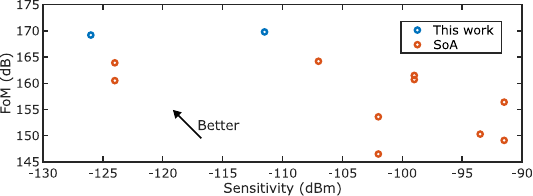}
\caption{Comparison to the state of the art in terms of the wake-up receiver FoM as a function of sensitivity.}
\label{fig: FoM_WU}
\end{figure}

Table~\ref{tab: WU_SoTA} shows the comparison of the proposed LoRa wake-up receiver with state-of-the-art long-range wake-up receivers. \firstRev{The two columns shown for this work correspond to the extreme SF values used in LoRaWAN, i.e., 7 and 12.} For a quantitative comparison, we use the following FoM:
\begin{equation}
\text{WU FoM} = -\text{Sens.} - 10\log(\text{Power}) - 10\log(\text{Latency}).
\end{equation}
The resulting FoM can be visualized as a function of the sensitivity in \figurename~\ref{fig: FoM_WU}. The proposed wake-up receiver achieves the best FoM, even though the total power consumption is not yet compatible with an IoT node. In terms of sensitivity, only the design in \cite{kim2023} achieves a good-enough sensitivity to match that of a main LoRa receiver. However, their design does not respect any wireless standard, which makes it more difficult to adopt in a network.

\section{Conclusions}
\label{sec: Conclusion}

In this paper, we present a \firstRev{programmable LoRa-compatible receiver with an integrated wake-up mode}. By dynamically exploiting the noise-power trade-off of both the LNA and the mixer, the receiver can change its NF between 4.8 and \qty{13.2}{dB} for a power consumption between 1.71 and \qty{4.57}{mW}. Power consumption savings of \qty{62.6}{\%} can be achieved at the receiver level at the expense of an 8.3-dB sensitivity degradation. \genRev{Under the same modulation parameters, the main receiver and the wake-up mode reach sensitivities of -138 and \qty{-126}{dBm} respectively}. To reach similar sensitivities in the wake-up mode and in the main receiver mode, this degradation should be compensated for. To do so, future works should leverage the repetition of upchirps in LoRa preambles. Preamble detection algorithms, as presented in~\cite{tapparelLoRaPreambleDetection2023,kang_lora_2023}, could be integrated inside the DBB with small logic overheads.

In wake-up mode, the receiver consumes between 1.46 and \qty{1.51}{mW}, corresponding to \qty{67}{\%} less than what the programmable consumes in the worst-case scenario. However, the power consumption is still high for an IoT node, as sub-mW consumption is needed to enable an always-on operation of the wake-up receiver for extending the lifetime of the node. The ADPLL is currently the power consumption hotspot of the WuRX. Combinations of technology scaling, specific PLL design, and injection locking techniques could be leveraged to address this issue. Moreover, duty-cycling of the WuRX could also be implemented.


A quantitative comparison with state-of-the-art \genRev{long-range receivers and wake-up receivers} shows a better FoM for the proposed design, thanks to its low sensitivity and latency. Further improvements are needed to enable its use on an IoT node, notably on the PLL or the use of duty-cycling, to make it more efficient.

\bibliographystyle{IEEEtran}
\bibliography{BREWER.bib}

\begin{IEEEbiography}[{\includegraphics[width=1in,height=1.25in,clip,keepaspectratio]{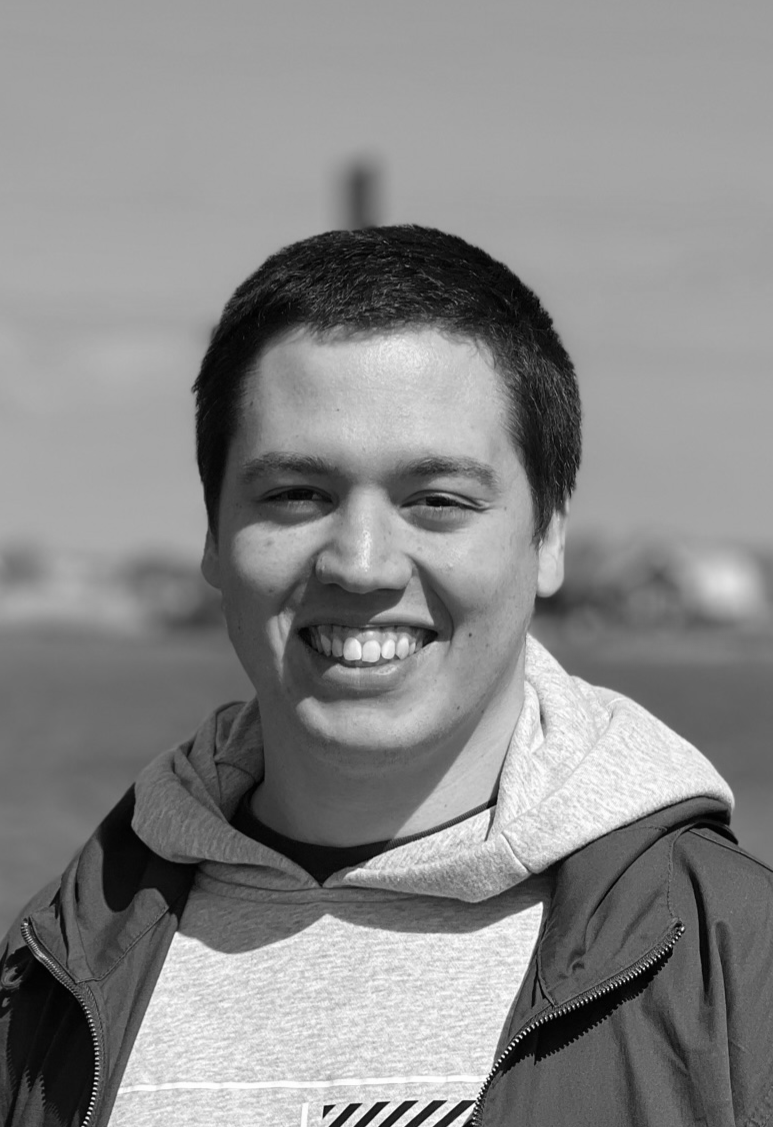}}]{Marco Gonzalez}
(Graduate Student Member, IEEE) received the M.Sc. degree in Electrical Engineering from the Université catholique de Louvain (UCLouvain), Louvain-la-Neuve, Belgium, in 2019, where he is currently pursuing a Ph.D. degree in microelectronics, under the supervision of Prof. David Bol. His current research includes the design of low-power analog, mixed-signal, and radio-frequency integrated circuits.
\end{IEEEbiography}
\vskip 0pt plus -1fil

\begin{IEEEbiography}[{\includegraphics[width=1in,height=1.25in,clip,keepaspectratio]{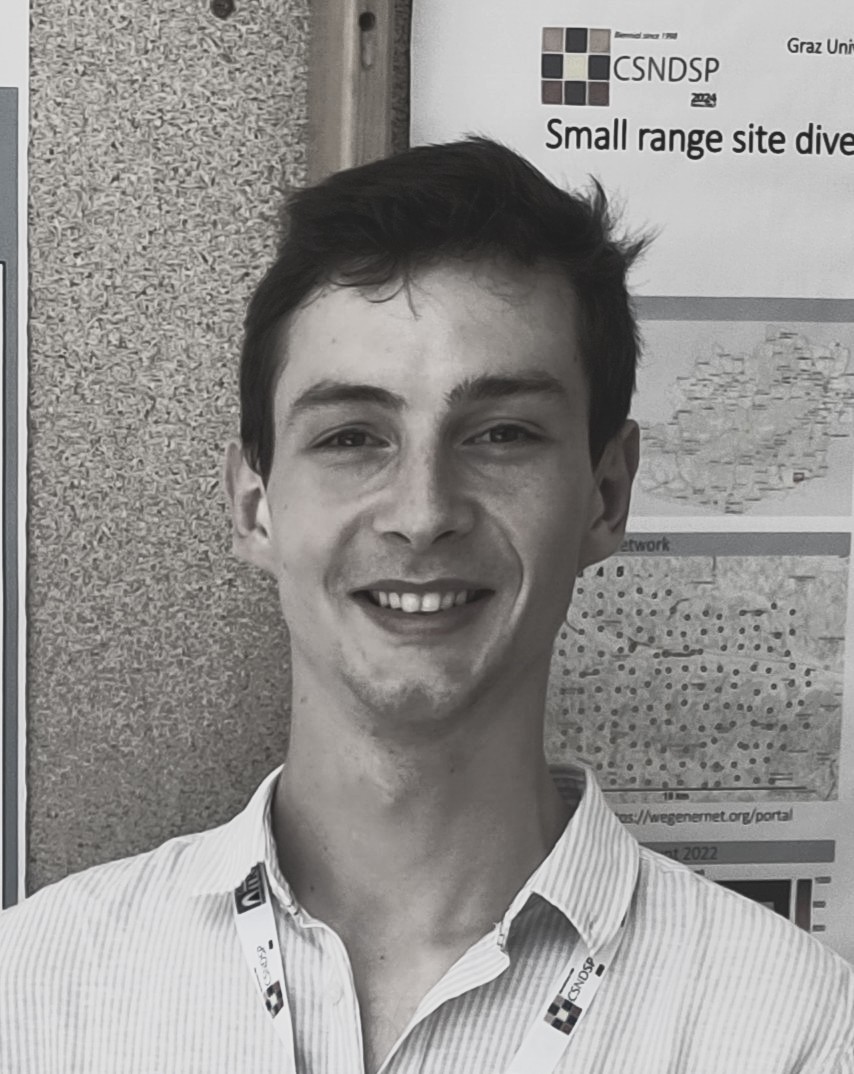}}]{Pol Maistriaux}
(Graduate Student Member, IEEE) received the M.Sc. degree in electrical engineering from the Université catholique de Louvain (UCLouvain), Louvain-la-Neuve, Belgium, in 2021, where
he is currently pursuing the Ph.D. degree with the Electronics Circuits and Systems Group, under the supervision of Prof. D. Bol. His current research interests include digital IC design and ultra-low-power receivers to enable large-scale mesh sensor networks for environmental monitoring applications.
\end{IEEEbiography}
\vskip 0pt plus -1fil

\begin{IEEEbiography}[{\includegraphics[width=1in,height=1.25in,clip,keepaspectratio]{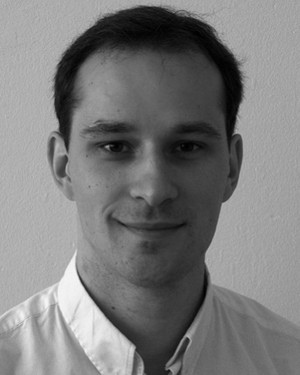}}]{Jérôme Louveaux}
(Member, IEEE) received the Ph.D. degree from the Université Catholique de Louvain (UCLouvain), Ottignies-Louvain-la-Neuve, Belgium, in 2000. From 2000 to 2001, he was a Visiting Scholar with Stanford University, Stanford, CA, USA. From 2004 to 2005, he was a Postdoctoral Researcher with Delft University of Technology, Delft, The Netherlands. Since 2006, he has been a Professor with the ICTEAM Institute, UCLouvain. His research interests include signal processing for digital communications, in particular, multicarrier modulations, xDSL systems, resource allocation, synchronization, and estimation.
\end{IEEEbiography}
\vskip 0pt plus -1fil

\begin{IEEEbiography}[{\includegraphics[width=1in,height=1.25in,clip,keepaspectratio]{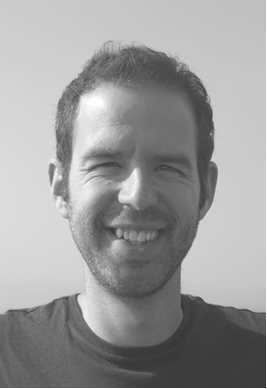}}]{David Bol}
(Senior Member, IEEE) received the M.Sc. degree in Electromechanical Engineering and the Ph.D degree in Engineering Science from Université catholique de Louvain (UCLouvain), Louvain-la-Neuve, Belgium, in 2004 and 2008, respectively. In 2005, he was a visiting Ph.D student at the CNM National Centre for Microelectronics, Sevilla, Spain, in advanced logic design. In 2009, he was a postdoctoral researcher at intoPIX, Louvain-la-Neuve, Belgium, in low-power design for JPEG2000 image processing. In 2010, he was a visiting postdoctoral researcher at the UC Berkeley Laboratory for Manufacturing and Sustainability, Berkeley, CA, in the life-cycle assessment of the semiconductor environmental impact. He is now an Associate Professor at UCLouvain. In 2015, he participated to the creation of e-peas semiconductors, Louvain-la-Neuve, Belgium. Prof. Bol leads the Electronic Circuits and Systems (ECS) group focused on ultra-low-power design of integrated circuits for the IoT and biomedical applications, including computing, power management, sensing, and wireless communications. He is engaged in a social-ecological transition in the field of ICT research. He co-teaches four M.S. courses in Electrical Engineering at UCLouvain on digital, analog, and mixed-signal integrated circuits and systems, as well as sensors, with two B.S. courses, including the course on Sustainable Development and Transition. On the private side, he pioneered the parental leave for male professors in his faculty to spend time connecting to nature with his family.
\end{IEEEbiography}

\end{document}